\documentclass{andp2012}
\usepackage[english]{babel}
\keywords{quantum state transfer, topological phases, Landau-Zener tunneling}
\title{Landau-Zener topological quantum state transfer}
\author[S: Longhi]{Stefano Longhi \inst{1,}\footnote{Corresponding author\quad E-mail:~\textsf{longhi@fisi.polimi.it}}}
\author[G: Girogi]{Gian Luca Giorgi \inst{2}}
\author[R: Zambrini]{Roberta Zambrini \inst{2}}
\address[1]{Dipartimento di Fisica, Politecnico di Milano and Istituto di Fotonica e Nanotecnologie del Consiglio Nazionale delle Ricerche, Piazza L. da Vinci 32, I-20133 Milano, Italy}
\address[2]{IFISC (UIB-CSIC), Instituto de Física Interdisciplinar y Sistemas Complejos (Universitat de les Illes Balears-Consejo Superior de
Investigaciones Científicas), UIB Campus, E-07122 Palma de Mallorca, Spain}
\shortauthors{S. Longhi, G.L.Giorgi, R. Zambrini,}
\begin{abstract}
Fast and robust quantum state transfer (QST) is a major requirement in quantum control and in scalable quantum information
processing. Topological protection has emerged as a promising route for the realization of QST robust against sizable imperfections in the network.
Here we present a scheme for robust QST of topologically protected edge states in a dimeric Su-Schrieffer-Heeger spin chain assisted by Landau-Zener tunneling. As compared to topological QST protocols based on Rabi flopping proposed in recent works, our method {is more advantageous} in terms of robustness against both diagonal and off-diagonal disorder in the chain, without a substantial increase of  the interaction time.
\end{abstract}
\shortabstract
\begin{document}
\maketitle
\section{Introduction}
Excitation transfer in classical and quantum networks is of major interest in different areas of science  and technology with a wealth of applications ranging from coherent control of chemical reactions \cite{r1} and efficient excitation transfer in organic molecules \cite{r2,r3,r4} to quantum state transfer (QST) and large-scale quantum information processing \cite{r5,r6,r7,r8,r9,r10,r11,r12,r12bis,r13,r14,r15,r16,r17,r18,r19,r19bis,r19tris}. For the latter application, quantum states need to be coherently and robustly transferred between distant nodes in a quantum network. In the past two decades, different schemes have been proposed to implement  QST in various physical systems. Examples include  probabilistic state transfer in a chain with uniform parameters \cite{r6}, perfect state transfer in time-independent chains with properly tailored hopping amplitudes \cite{r10,r11,r20,r21,r22}, state transfer using externally applied time-dependent control fields \cite{r14,r15,r19bis}, Rabi flopping of nearly-resonant edge states \cite{r17}, adiabatic, superadiabatic and topologically-protected QST schemes \cite{r12bis,r23uff,r23,r24,r25,r25bis,r25tris,r25quatris,r26,r27,r27bis,r28}. A major requirement of QST protocols is to be robust against sizable imperfections in the network. To this regard, topological QST methods, where a quantum state can be stored and transmitted in a topologically-protected manner, have attracted great interest in the past few years owing to the opportunity to harvest topological phenomena for guiding and transmitting quantum information reliably \cite{r23,r24,r25,r25bis,r26,r27,r27bis,r28}.  The Su-Schrieffer-Heeger (SSH) model, originally introduced to describe transport properties of the conductive polyacetylene \cite{r29}, provides perhaps the most basic model system supporting topological excitations protected by chiral symmetry that is a promising setting for the realization of topological QST \cite{r12bis,r23,r25bis,r27,r27bis,r28}. In the SSH dimeric chain, two distinct QST protocols have been suggested, depending on whether the chain comprises an odd or even number of sites. For a SSH chain with an odd number of sites, i.e. with half integer dimers, there is only one edge state, which is localized either at the left or right edges of the chain depending on whether the intra- to inter-hopping rate ratio $r=t_2/t_1$ is larger or smaller than one. By adiabatically varying the ratio $r$, from below to above one, QST is realized by pumping the localized state from one edge to the other one (Thouless pumping) \cite{r12bis,r25,r28}. Since the edge state is topologically protected against perturbations that do not break chiral symmetry, this QST protocol shows partial protection against structural imperfections of the hopping amplitudes in the chain (off-diagonal disorder). However, it remains sensitive to on-diagonal disorder, i.e. disorder of site energies. For a SSH chain with an even number of sites, i.e. with an integer  number of dimers, in the non-trivial topological phase $r<1$ there are two edge modes. For finite chains, the two edge modes hybridize and undergo Rabi-like oscillations, which can be exploited to realize QST between the two edge sites of the chain \cite{r25bis,r27,r27bis}. For static chains, the time required to achieve QST with a  high fidelity turns out to be extremely long \cite{r25bis,r27bis}, which is undesirable owing to decoherence effects. Moreover, a careful timing of the interaction is required, preventing the possibility to delay the transfer process on demand. Recently, a protocol has been suggested to shorten the transit time, where the ratio $r$ of hopping rates is adiabatically varied to confine $(r \simeq 0$), delocalize and interfere ($r \simeq 1$), and then relocalize again ($r \simeq 0$) the two edge states \cite{r27}. However, the time for QST is affected by structural disorder in the chain, even though the disorder is only off-diagonal and does not break the chiral symmetry of the underlying Hamiltonian. Hence, the intrinsic robustness of the topological edge states is not fully exploited in such a QST scheme. \\
In this article we suggest a different route for topological QST in a SSH chain which is robust against both off-diagonal and on-diagonal structural disorder in the chain. We consider a SSH chain with an integer number of dimers \cite{r25bis,r27,r27bis} and realize QST between the two topological edge modes via a Landau-Zener (rather than Rabi flopping) transition, which is robust against both off- and on-diagonal disorder of the chain. As compared to QST based on Rabi flopping of adiabatically-deformed topological edge states \cite{r27}, the increase in transfer time is minimal while high fidelity is observed even for a moderate-to-strong disorder in the chain.
\begin{figure}
  \includegraphics[width=\columnwidth]{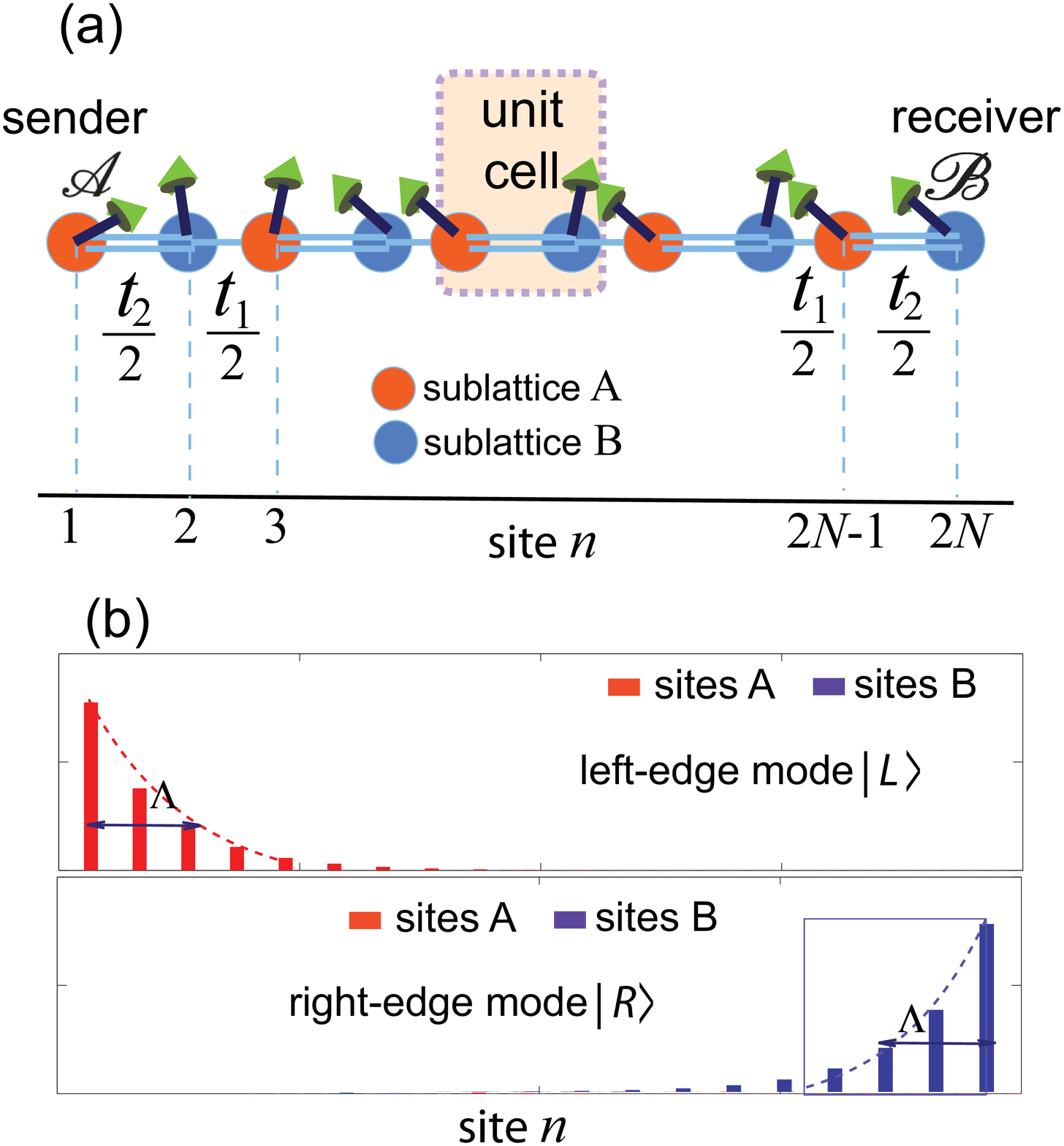}%
  \caption{\label{<label name>}\col 
   (a) Schematic of a dimerized spin-1/2 chain comprising $N$ dimers for topological QST. Sender $\mathcal{A}$ and receiver $\mathcal{B}$ are the edge sites of the chain. (b) Amplitude distribution of left ($L$) and right ($R$) edge states of the SSH chain in the non-trivial topological phase $t_2<t_1$. $L$ and $R$ edge states occupy only the sites of sublattice A and B of the chain, respectively. The localization length $\Lambda$ of the edge states is determined  by the ratio $r=t_2/t_1$, with strong localization in the $r \rightarrow 0$ limit (flat band limit) and delocalization in the $r \rightarrow 1$ limit (gap closing limit).}
\end{figure}

 \section{Quantum State Transfer in a dimerized spin chain}
As a paradigmatic model of QST, we consider the transfer of a single qubit in spin-1/2 chain systems \cite{r5,r6}, however different setups could be envisaged, such as superconducting qubit chains \cite{r19tris,r28} and optical waveguide lattices \cite{r20,r21,r22,r23,uffa1}. {{ In photonic systems, topologically-protected light guiding has been demonstrated in several experiments \cite{Alex1,Alex2,Alex3}}}, and adiabatic transport of topological edge states via Thouless pumping has been reported using either classical or quantum light \cite{r23,r25,r34bis}.

 Let us assume a dimerized spin chain \cite{r31} comprising $N$ dimers with spins coupled through the nearest-neighbor XX model with alternating coupling strengths $t_1/2$ and $t_2/2$ [Fig.1(a)]. Staggered magnetic fields, with amplitudes $\delta/2$ and $-\delta/2$, are applied at sublattices A and B of the spin chain. The Hamiltonian of the system reads \cite{r5,r6,r31}
\begin{equation}
\hat{H}=\sum_{n=1}^{2N-1}J_n ( \sigma_n^x\sigma_{n+1}^{x}+ \sigma_n^y\sigma_{n+1}^{y})+\sum_{n=1}^{2N} h_n \sigma_n^z
\end{equation} 
 where $J_n=t_1/2$ for $n$ even, $J_n=t_2/2$ for $n$ odd, and $h_n=-(-1)^n \delta/2$. In the standard protocol of one-qubit QST \cite{r5},
 the initial state, encoded on the left-edge sender spin ${\mathcal A}$, is assumed to be given by
$| \psi (0) \rangle = \alpha |0 \rangle_z + \beta |1 \rangle_z$, with $|\alpha|^2+|\beta|^2=1$ ($|0\rangle_z$  and $|1\rangle_z$ denote the spin-up and
-down states along the $z$ axis, respectively), whereas the other sites of the chain
are prepared with all spins up.  The
efficiency of the state transfer to the right-edge receiver spin $\mathcal{B}$ at time $t$ is quantified by the fidelity $\mathcal{F}(t)$, which equals 1 for a perfect
transfer. In order to evaluate the channel quality independently
of the specific input state, one usually introduces the average fidelity $\bar{\mathcal{F}}(t )$, which is obtained from $\mathcal{F}(t)$
after averaging over all possible pure input states of the qubit. The average fidelity reads \cite{r3,r18}
 \begin{equation}
 \bar{\mathcal{F}}(t )=\frac{1}{2}+\frac{1}{3}|f(t)|+ \frac{1}{6} | f(t)|^2
 \end{equation}
 where $f(t)$ is the transition amplitude of a
spin excitation from the left to the right edge sites of the chain. Clearly, a high average fidelity is achieved whenever the excitation transfer probability $|f(t)|^2$ is close as much as possible to one.
 Since the dynamics occurs in the subspace of single excitation sector, $f(t)$ can be calculated 
from the hopping dynamics of a single spinless particle along a tight-binding chain with alternating hopping rates $t_1$,$t_2$ and site potentials $\pm \delta$ in the two sublattices A and B \cite{r6,r17,r18,r31}. 
 After writing $|\psi(t) \rangle=\sum_{n=1}^{2N} c_n(t) |n \rangle$ for the vector state of the spineless particle hopping on the chain, the evolution equations of the occupation amplitudes
$c_n$ at the various sites $|n \rangle$ of the chain, as obtained from the  single-particle Schr\"odinger equation, read
 \begin{equation}
 i \frac{dc_n}{dt}=\sum_{m=1}^{2N}\mathcal{H}_{n,m}c_m
 \end{equation}
 ($n=1,2,...,2N$) where the $2N \times 2N$ matrix Hamiltonian $\mathcal{H}$ is the Rice-Mele Hamiltonian \cite{r37bis}, given by
 \begin{equation}
 \mathcal{H}= \left(
 \begin{array}{cccccccccc}
 \delta & t_2 & 0 & 0 & 0 & ...& 0 & 0 & 0  &0\\
 t_2 & -\delta & t_1 & 0 & 0 & ... & 0 & 0 & 0 & 0 \\
 0 & t_1 & \delta & t_2 & 0 &  ... & 0 & 0 & 0  & 0\\
 ... & ... & ...& ...& ... & ...& ...& ...& ...& ...  \\
 0 & 0& 0 & 0& 0 & ...    & 0 & t_1 &\delta & t_2 \\
 0 & 0& 0 & 0& 0 & ...    & 0 & 0 & t_2 &- \delta  \\
 \end{array}
 \right).
 \end{equation}
 Note that $\mathcal{H}$ reduces to the SSH model in the $\delta=0$ limit.
The single-particle transfer excitation amplitude $f(t)$, that determines the average fidelity according to Eq.(2), is given by $f(t)=c_{2N}(t)$, where $c_{2N}(t)$ is the solution to Eq.(3) with the initial condition $c_n(0)=\delta_{n,1}$.\\ 
Let us first briefly review the QST protocols based on Rabi flopping of left ($L$) and right ($R$) topological edge states, recently introduced in Refs.\cite{r27,r27bis}. In such protocols, one assumes $\delta=0$ (no local magnetic fields) and the non-trivial topological phase $ r \equiv t_2/t_1<1$ of the SSH chain, which ensures the existence of topological edge states. The state transfer arises because of hybridization of the $L$ and $R$ edge states in the finite chain, which occupy the A and B sublattices, respectively [Fig.1(b)]. They are defined by
 \begin{eqnarray}
 |L \rangle & = \mathcal{N} & \sum_{n=1,3,5,...,2N-1} (-t_2/t_1)^{(n-1)/2} | n \rangle \\
 |R \rangle & = \mathcal{N }& \sum_{n=2,4,6,...,2N} (-t_2/t_1)^{(N-n/2)} | n \rangle.
 \end{eqnarray}
 where  
 \[ \mathcal{N}=\sqrt{\frac{1}{\sum_{n=0}^{N-1} (t_2/t_1)^{2n}}}= \sqrt{\frac{r^2-1}{r^{2N}-1}}\]
  is the normalization factor.
 Strictly, the $L$ and $R$ edge states defined by Eqs.(5) and (6) are exact eigenmodes of the Hamiltonian $\mathcal{H}$ only for semi-infinite chains, i.e. when the chain is truncated only at the left or right edges, respectively. In this limiting case, $| L \rangle$ and $| R \rangle$ are zero-energy degenerate modes with topological protection for off-diagonal disorder (hopping rate disorder) that does not close the gap. Both edge states are exponentially localized  with a localization length (measured in units of lattice period) given by
 \begin{equation}
 \Lambda \sim  \frac{1}{2 \log (t_1/t_2)}.
 \end{equation}
  Note that $\Lambda$ shrinks to zero as $t_2 / t_1 \rightarrow 0$ (flat band limit), while $\Lambda$ diverges as $t_2/t_1 \rightarrow 1$ (gap closing limit). Thus, the two edge states are well overlapped with the sender $\mathcal{A}$ and receiver $\mathcal{B}$ sites provided that $t_2 / t_1 \ll 1$. For a finite chain of $N$ dimers the $L$ and $R$ modes hybridize and the zero-energy degeneracy is lifted. In fact, in the subspace described by the vectors $|L \rangle$ and $|R \rangle$ defined by Eqs.(5) and (6), after expanding the vector state as 
  \begin{equation}
  | \psi(t) \rangle= a_L(t) |L \rangle+ a_R(t) |R \rangle
  \end{equation}
   the reduced two-state dynamics of amplitudes $a_{R,L}(t)$ reads (see Appendix A)
\begin{eqnarray}
i \frac{da_L}{dt} & = & \kappa a_R \\
i \frac{da_R}{dt} & = & \kappa a_L 
\end{eqnarray}
where we have set
\begin{equation}
\kappa \equiv  \frac{t_1 \left( t_2/t_1\right)^{N} \left[ (t_2/t_1)^2-1\right]}{(t_2/t_1)^{2N}-1}.
\end{equation}
 Equations (9) and (10) show that in the finite chain the two edge state eigenvectors of the Hamiltonian $\mathcal{H}$ are approximately given by the odd/even superpositions $(|L \rangle \pm | R \rangle) / \sqrt{2}$ of $L$ and $R$ states, with eigen-energies $ \pm \kappa$. Interestingly, if at time $t=0$ the particle is prepared in state $L$, with strong overlap with the sender state $\mathcal{A}$ and $L$, i.e. assuming $t_2 / t_1 \ll 1$ and $a_L(0)=1$, $a_R(0)=0$, at time $t=T$ with
 \begin{equation}
 T = \frac{\pi}{2 \kappa}
 \end{equation}
 \begin{figure}
  \includegraphics[width=\columnwidth]{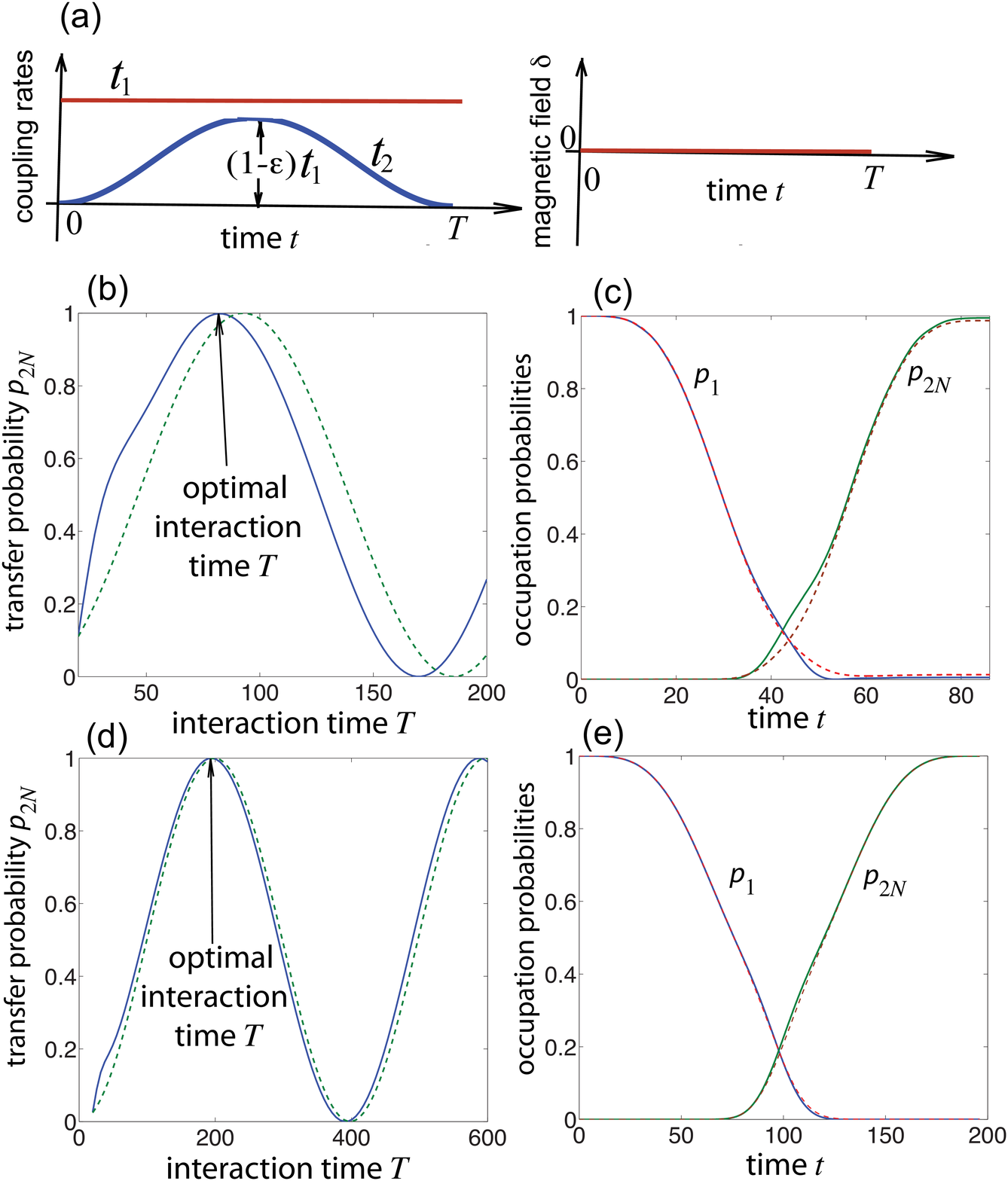}%
  \caption{\label{<label name>}\col 
   Topological QST based on Rabi flopping with adiabatic deformation of edge states (adiabatic Rabi protocol \cite{r27}). (a) Temporal behavior of the coupling constants $t_1$ and $t_2$ (left panel) and of local magnetic field $\delta$ (right panel). Note that in the adiabatic Rabi flopping scheme of QST the local staggered magnetic field is zero. (b) Behavior of the transfer excitation probability $p_{2N}$ versus interaction time $T$ for the Rabi protocol (13) with $\epsilon=0.1$ {and N=10}. Solid and dashed curves refer to exact numerical results and approximate two-level state analysis. (c) Detailed temporal evolution of occupation probabilities of sender ($p_1$) and receiver ($p_{2N}$) sites versus time $t$ for the optimal interaction time $T=86$. (d,e): Same as (b,c), but for $\epsilon=0.2$. In (e) the interaction time is $T=196$.}
\end{figure}
  one has $a_L(T)=0$ and $a_R(T)=-i$, indicating excitation transfer from $L$ to $R$ edge states (Rabi flopping). This is basically the transfer method considered in Refs.\cite{r25bis,r27bis}. The main limitation of this transfer scheme is that, in order to achieve transfer from $\mathcal{A}$ to $\mathcal{B}$ with high fidelity, the ratio  $ r =t_2 /t_1$ should be chosen as much as small possible, corresponding to an extremely long transit time $T$ according to Eqs.(11) and (12). A variant of the Rabi-flopping QST scheme, which considerably reduces the transit time $T$, has been recently proposed in Ref.\cite{r27}. The main idea is to adiabatically change the localization length of the edge states $L$ and $R$ by varying in time the ratio $r=t_2/t_1$, from zero at $t=0$ to a value $r=1-\epsilon$ at $t=T/2$ and then back to zero at $t=T$. For example, one can assume the adiabatic transfer protocol \cite{r27}
\begin{equation}
t_1=1 \; , \;\; t_2=\frac{1-\epsilon}{2} \left[1-\cos(2 \pi t /T) \right] \; , \;\; \delta=0
\end{equation}  
  as shown in Fig.2(a). In this case, at $t=0,T$, where $r=0$, the $L$ and $R$ edge states are tightly confined and exactly coincide with the sender ($\mathcal{A}$) and receiver ($\mathcal{B}$) edge sites of the chain, respectively, while at intermediate times the two states $L$ and $R$ are delocalized and they can undergo Rabi flopping in a short time (since $\kappa$ takes a non-negligible value). The parameter $\epsilon$ ($0< \epsilon <1$) entering in Eq.(13) determines the band gap of the SSH lattice at time $t=T/2$, with $\epsilon \rightarrow 0$ corresponding to a closing gap and $\epsilon \rightarrow 1$ to a flat band. In the adiabatic regime, a rough estimation of the minimum interaction time $T$ required to realize QST is obtained from the $^{\prime}$area theorem$^{\prime}$
  \begin{equation}
  \int_0^T \kappa(t) dt= \pi/2
  \end{equation}
  An example of QST based on the adiabatic Rabi protocol is shown in Figs.2(b-e). Figure 2(b) shows the behavior of the excitation transfer probability $p_{2N}(T) \equiv |f(T)|^2=|c_{2N}(T)|^2$ versus interaction time $T$ as obtained by  numerical solution of the Schr\"odinger equation (3) (solid curve) with the initial condition $c_n(0)=\delta_{n,1}$ for a chain comprising $N=10$ dimers and assuming  $\epsilon=0.1$ in Eq.(13). The dashed curve in figure shows the corresponding behavior of the transfer probability  $p_{2N}(T)$ as obtained by the approximate two-level model. The minimum optimal transfer time is obtained at $T \simeq 86$, corresponding roughly to the condition (14) (area theorem). A detailed behavior of the occupation probabilities of sender ($p_1(t)=|c_1(t)|^2$) and receiver ($p_{2N}(t)=|c_{2N}(t)|^2$) sites in the chain, for the optimal interaction time $T=86$, is shown in Fig.2(c). The main discrepancy between the exact and approximate two-level model results observed in Figs.2(b) and (c) can be mainly ascribed to the \small value of $\epsilon$ chosen in the simulations, corresponding to a small gap near $t=T/2$ and rather delocalized $L$ and $R$ states. At larger values of $\epsilon$ the two-state approximation clearly provides a more accurate description of the dynamics [see for example the results shown in Figs.2(d) and (e), where $\epsilon=0.2$], however this would require a longer interaction time.
  \par
 The adiabatic Rabi flopping scheme enables to greatly reduce the interaction time as compared to a static model, thus avoiding decoherence effects. However, this method is sensitive not only to diagonal (on-site) disorder in the chain, but also to disorder in the coupling constants (off-diagonal disorder), in spite of the topological nature of edge states (see Sec.4 below). The main reason thereof is that, since the coupling $\kappa$ of $L$ and $R$ edge states is an integral overlap  of $L$ and $R$ modes (see Appendix A),  its value [and thus the optimal transfer time $T$ satisfying the area theorem (14)] is sensitive to off-diagonal disorder. In other words, while off-diagonal disorder does not break chiral symmetry of the lattice, thus protecting the zero-energy value of edge modes in the large (thermodynamic) $N$ limit, in the finite chain the disorder modifies the profile of edge states and thus their energy splitting $2 \kappa$. Therefore, the optimal interaction time $T$ is sensitive to disorder in the chain, requiring a careful timing of the interaction to avoid degradation of fidelity. 
  
  \section{Landau-Zener topological quantum state transfer}
  In two-state systems, it is well known that adiabatic Landau-Zener (LZ) tunneling is a much more robust method than Rabi flopping to realize excitation transfer. The LZ model is one of the
most widely used two-state approximations in resonance physics and found broad applications in different areas of science, such as in atomic and molecular physics, quantum optics, chemical physics, etc. (see, e.g., \cite{r30} and references therein). In quantum control and quantum information science, several works {{in different experimental settings}} pointed out that LZ tunneling may provide
a simple and effective solution for the realization of high fidelity quantum state control without the need for precise timing \cite{r30a,r30b,r30c,r30d,r30e,r30f,r32}. {{Since the earlier experimental demonstrations of LZ interferometry in strongly-driven superconducting qubits \cite{palle1,palle2}, adiabatic rapid passage techniques are nowadays routinely realized in superconducting qubit systems. For example, interference in a superconducting qubit under periodic latching modulation, in which the level separation is switched abruptly between two values
and is kept constant otherwise, has been demonstrated in \cite{r19tris}, whereas fast and high-fidelity perfect quantum state transfer in a superconducting qubit chain with parametrically tunable couplings has been recently reported  in \cite{palle3}}}. Such previous studies suggest us that LZ tunneling of topological edge states in the SSH chain, besides of avoiding the timing problem of Rabi-like QST methods, could provide a viable route for high-fidelity QST which is robust against both diagonal and off-diagonal disorder of the chain \cite{r32,r33}. The main idea is to add a staggered local magnetic field  $\delta$, of opposite sign in the two sublattices A and B of the spin chain, which is linearly and slowly  ramped in time so as to realize LZ tunneling between the two edge states when they are delocalized in the chain. A schematic of the topological QST protocol based on LZ transition is shown in Fig.3(a) and corresponds to the following time-dependent parameters in the Rice-Mele Hamiltonian (4) [compare with Eq.(13)]
\begin{eqnarray}
t_1 & = & 1 \nonumber \\
t_2 & = & \left\{
\begin{array}{ll}
\frac{1-\epsilon}{2} \left[1-\cos( \pi t /\tau) \right]  & 0<t<\tau \\
1- \epsilon & \tau<t<\tau+\tau_Z \\
\frac{1-\epsilon}{2} \left[1-\cos( \pi (t-\tau_Z) /\tau) \right]  & \tau+\tau_Z<t<T 
\end{array}
\right. \\
\delta & = & \left\{
\begin{array}{ll}
\delta_0 & 0<t< \tau \\
\delta_0-\alpha(t-\tau)/2 & \tau<t<\tau_z+\tau \\
-\delta_0 & \tau+\tau_Z<t<T
\end{array}
\right. \nonumber
\end{eqnarray}  
\begin{figure}
  \includegraphics[width=\columnwidth]{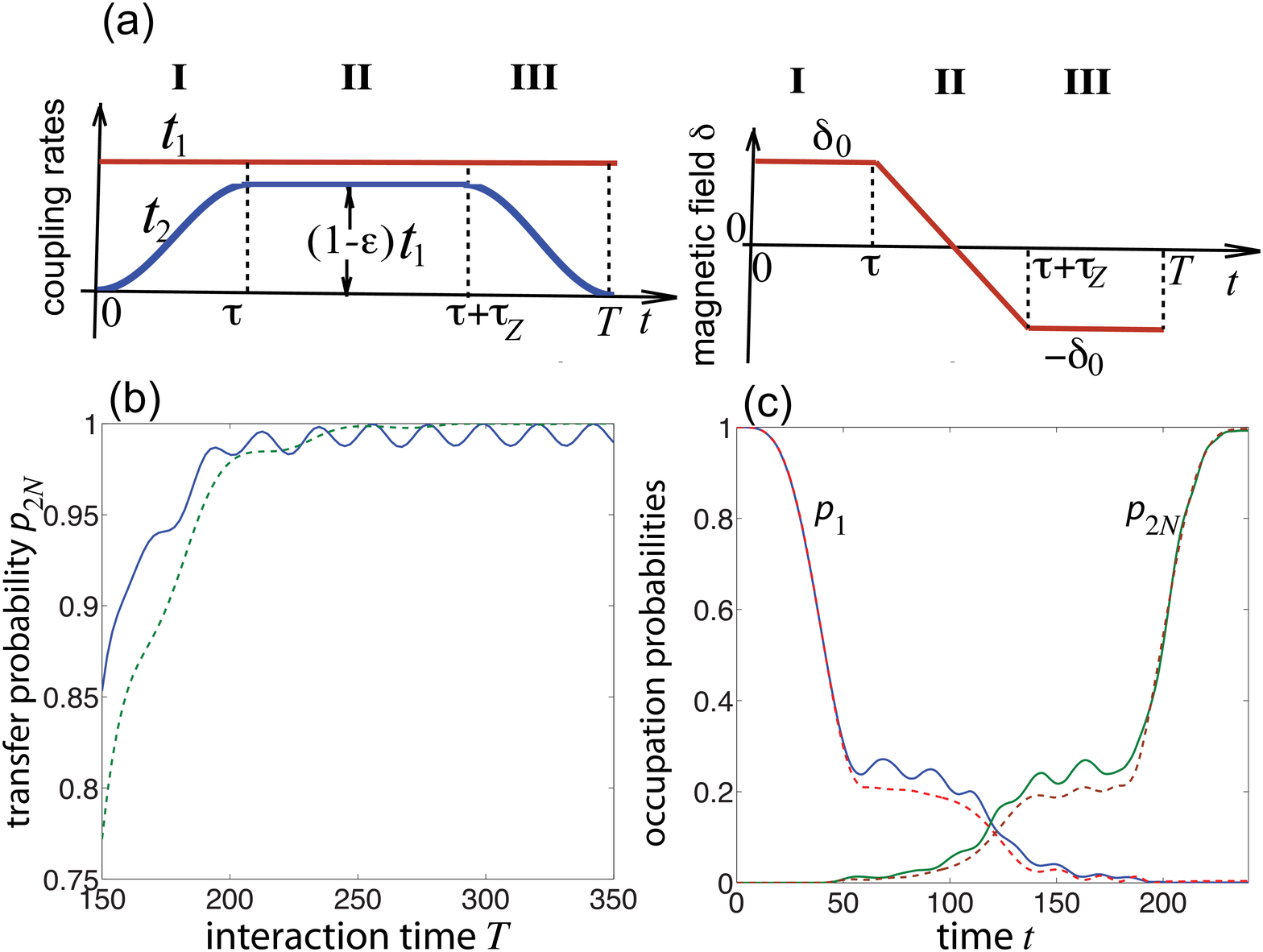}%
  \caption{\label{<label name>}\col 
   Landau-Zener topological QST. (a) Temporal behavior of the coupling constants $t_1$ and $t_2$ (left panel) and of local staggered magnetic field $\delta$ (right panel). The QST protocol comprises three stages: in stages I and III the $L$ and $R$ edge states are adiabatically delocalized (stage I) and relocalized (stage III), like in the adiabatic Rabi flopping scheme of Fig.2(a), however interaction is forbidden by the staggered field $\delta_0$. In stage II LZ tunneling is realized by sweeping the magnetic field from $\delta_0$ to $-\delta_0$ in a time interval $\tau_Z$. (b) Behavior of the transfer excitation probability $p_{2N}$ versus interaction time $T=2 \tau+\tau_Z$ for the LZ protocol (15) and for parameter values $\epsilon=0.1$, $\delta_0=0.2$, $\tau=60$. The numbers of dimers in the chain is $N=10$. Solid and dashed curves refer to exact numerical results and approximate two-level state analysis. (c) Detailed temporal evolution of occupation probabilities of sender ($p_1$) and receiver ($p_{2N}$) sites versus time $t$ for the interaction time $T=240$.}
\end{figure}
\begin{figure}
  \includegraphics[width=\columnwidth]{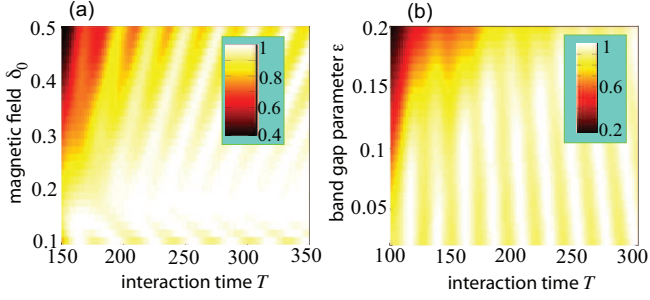}%
  \caption{\label{<label name>}\col 
   (a) Pseudocolor map showing the dependence of the transfer probability $p_{2N}$ in the $(T,\delta_0)$ plane for the LZ QST protocol of Fig.3(a) and for parameter values $\epsilon=0.1$ and $\tau=60$.
   (b) Pseudocolor map showing the dependence of the transfer probability $p_{2N}$ in the $(T, \epsilon)$ plane for parameter values $\delta_0=0.2$, $\tau=40$ and $N=10$.}
\end{figure}
where $T=2 \tau+\tau_Z$ is the interaction time and $\alpha=4 \delta_0 / \tau_Z$ is the temporal gradient of the local magnetic field. Note that the transfer scheme comprises three stages: in the first stage I (time duration $\tau$), the two edge states are adiabatically delocalized as in the adiabatic Rabi scheme of Fig.2(a), however the applied local magnetic field $\delta_0$ splits the energies of the two edge states far apart so that they do not interact. In the second stage II (duration $\tau_Z$) the magnetic field is linearly decreased in time till to vanish and reverse sign, while the ratio $r=t_2/t_1$ is kept constant at a value close to one:  in this time interval LZ tunneling  between the delocalized $L$ and $R$ states occurs. Finally, in the third step III (time duration $\tau$) the two edge states are adiabatically re-localized at the edge sites. In the spirit of the two-level approximation, the excitation transfer between the sender and receiver edge sites of the chain is described by the coupled equations (see Appendix A)
\begin{eqnarray}
i \frac{da_L}{dt} & = &   \delta(t) a_L+\kappa(t) a_R \\
i \frac{da_R}{dt} & = &  - \delta(t) a_R+\kappa(t) a_L 
\end{eqnarray}
where $\kappa=\kappa(t)$ is given by Eq.(11) and the time dependence of $\delta$ and $t_2$ is defined by Eq.(15). We require $\delta_0 > \sim \kappa$ so that the two edge states are decoupled in stages I and III. Under such an assumption, the transition probability is given by the well-known Landau-Zener relation \cite{r30} $p_{2N} \simeq 1-\exp(-2 \pi \Gamma)$, with $\Gamma= \kappa^2/ \alpha$. Hence, a high excitation transfer is realized provided that $\Gamma > \sim 1$, i.e. 
\begin{equation}
\tau_Z > \sim \frac{4 \delta_0}{\kappa^2}
\end{equation}
\begin{figure}
  \includegraphics[width=\columnwidth]{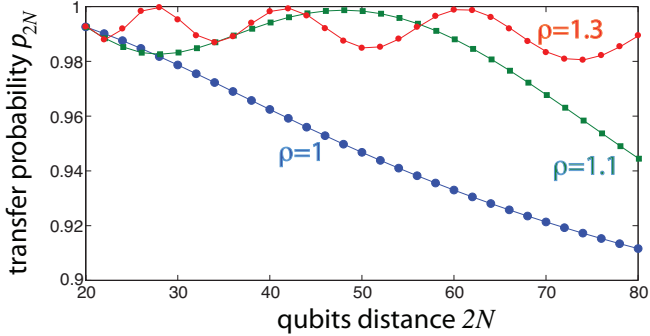}%
  \caption{\label{<label name>}\col 
   {{Behavior of the transfer probability $p_{2N}$ versus the distance $2N$ between the quits for the LZ protocol of Fig.3(a). The following dependence of parameters $\epsilon$, $\delta_0$, $\tau$ and $T$ on $N$ is assumed: $\epsilon=1/N$, $\delta_0=2/N$, $\tau=60 \times (N/10)^{\rho}$ and $T=240 \times (N/10)^{\rho}$. Note that for $N=10$ the parameter values correspond to the simulation shown in Fig.3(c).}}}
\end{figure} 
with $\delta_0 > \sim \kappa$. As an example, Fig.3(b) shows the numerically-computed behavior of the transfer probability $p_{2N}$ for parameter values {{$N=10$}}, $\epsilon=0.1$, $\tau=60$, $\delta_0=0.2$ and for increasing values of the LZ time $\tau_Z$, i.e. of the interaction time $T=2 \tau+\tau_Z$. Solid and dashed curves in the figure refer to the full numerical simulations of the Schr\"odinger equation and to the approximate two-level model, respectively. Clearly, for a sufficiently long LZ time $\tau_Z$ [$\tau_Z > \sim 80$ in the simulation of Fig.3(b)], efficient excitation transfer is realized, which becomes largely insensitive to a change of $\tau_Z$, thus indicating that -- unlike in the Rabi flopping scheme-- precise timing of interaction is not required in the topological LZ QST protocol. An example of the detailed behavior of the occupation probabilities at sender ($p_1(t)=|c_1(t)|^2$) and receiver ($p_{2N}(t)=|c_{2N}(t)|^2$) sites, for a transit time $T=240$, is shown in Fig.3(c). Note that, as compared to the Rabi-flopping scheme of Fig.2, the LZ adiabatic scheme requires a longer interaction time (due to the additional LZ time $\tau_Z$), however the increase of transfer time $T$ is moderate (less than one order of magnitude). Parameter optimization to obtain a high-fidelity transfer in in the shortest possible interaction time $T$ would require full numerical simulations to scan the entire 4-dimensional parameter space $\epsilon$, $\delta_0$, $\tau$ and $\tau_Z$, with $T=2 \tau+\tau_Z$. This is a rather cumbersome task which goes beyond the scope of the present work. However, extended numerical simulations in reduced 2-dimensional space indicate that there exist wide range of parameters where high values of transfer probability ($p_{2N}$ larger than $0.95$) can be  achieved with an interaction time $T$ few times larger than the one typically required in the adiabatic Rabi scheme of Ref.\cite{r27}. As an example, Figs.4(a) and (b) show numerically-computed maps of the transfer probability $p_{2N}$  in the $(\delta_0,T)$ and $(\epsilon,T)$ planes, respectively, for fixed values of other parameters. { {The results shown in the figures refer to the exact numerical simulations of the Schr\"odinger equation (3), i.e. beyond the two-level approximation}}. The broad white areas in the plots, corresponding to a transfer probability larger than $\sim 0.95$, clearly indicate that high-fidelity QST can be achieved without any precise fine tuning of parameter values. {{Finally, let us discuss about the scalability of the adiabatic LZ protocol with separation between the  two qubits, i.e. number $2N$ of sites in the chain. Like in the adiabatic Rabi protocol \cite{r27}, the interaction time $T$ required to realize state transfer with a high fidelity is ultimately limited by the finite propagation speed of excitation in the chain, expressed by the Lieb-Robinson bound \cite{Robinson}, and by the adiabaticity criterion to avoid losses into the bulk states of the SSH lattice. In practice, in optimized protocols the dependence of transfer time $T$ on lattice sites $2N$ scales with the algebraic law $T \sim (2N)^ \rho$ with $\rho \geq 1$ \cite{r27}, the lowest value $\rho=1$ corresponding to the Lieb-Robinson bound \cite{r27}. Figure 5 shows the numerically-computed behavior of the transfer probability $p_{2N}$ versus the quits distance $2N$ for the three values of the exponent $\rho=1$,  1,1 and 1.3. Parameter values are as in Fig.3, expect that at each value of $2N$ all the time constants are scaled by the factor $\sim (2N)^{\rho}$ while $\epsilon$ and $\delta_0$ are scaled by the factor $\sim 1/N$. The results clearly indicate that, for an interaction time $T$ that increases slightly more than linear with the size $2N$ of the chain (curve with $\rho=1.3$), the probability transfer remains larger than $ 98 \%$ over the entire range from $2N=20$ to $2N=80$.}
\begin{figure}
  \includegraphics[width=\columnwidth]{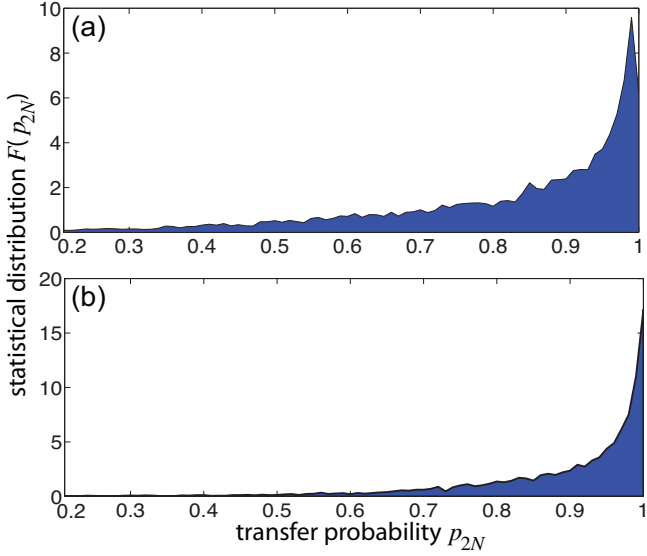}%
  \caption{\label{<label name>}\col 
   Effect of disorder on the transfer probability $p_{2N}$ in the adiabatic Rabi scheme of Fig.2(a) for parameter values $\epsilon=0.1$ and $T=86$. {{The number of lattice sites in $N=10$.}} (a) Diagonal disorder (disorder strength $\delta E=0.2$). (b) Off-diagonal disorder of inter-dimer hopping rate $t_1$ (disorder strength $\sigma=0.2$). The statistical distribution $F(p_{2N})$ of $p_{2N}$ is obtained assuming 10000 realizations of disorder.}
\end{figure} 
\begin{figure}
  \includegraphics[width=\columnwidth]{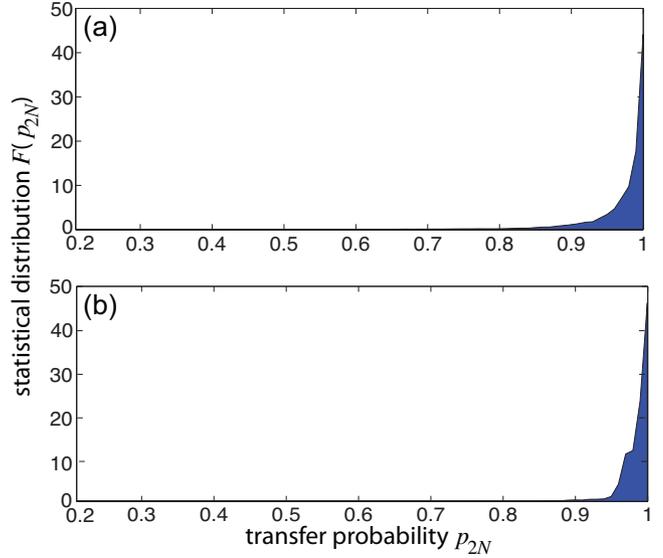}%
  \caption{\label{<label name>}\col 
   Effect of disorder on the transfer probability $p_{2N}$ in the LZ scheme of Fig.3(a) for parameter values $\epsilon=0.1$, $\tau_Z=120$, $\delta_0=0.2$ and $T=240$. {{The number of lattice sites is $N=10$.}} (a) Diagonal disorder (disorder strength $\delta E=0.2$). (b) Off-diagonal disorder of inter-dimer hopping rate $t_1$ (disorder strength $\sigma=0.2$). The statistical distribution $F(p_{2N})$ of $p_{2N}$ is obtained assuming 10000 realizations of disorder.}
\end{figure} 
\begin{figure}
  \includegraphics[width=\columnwidth]{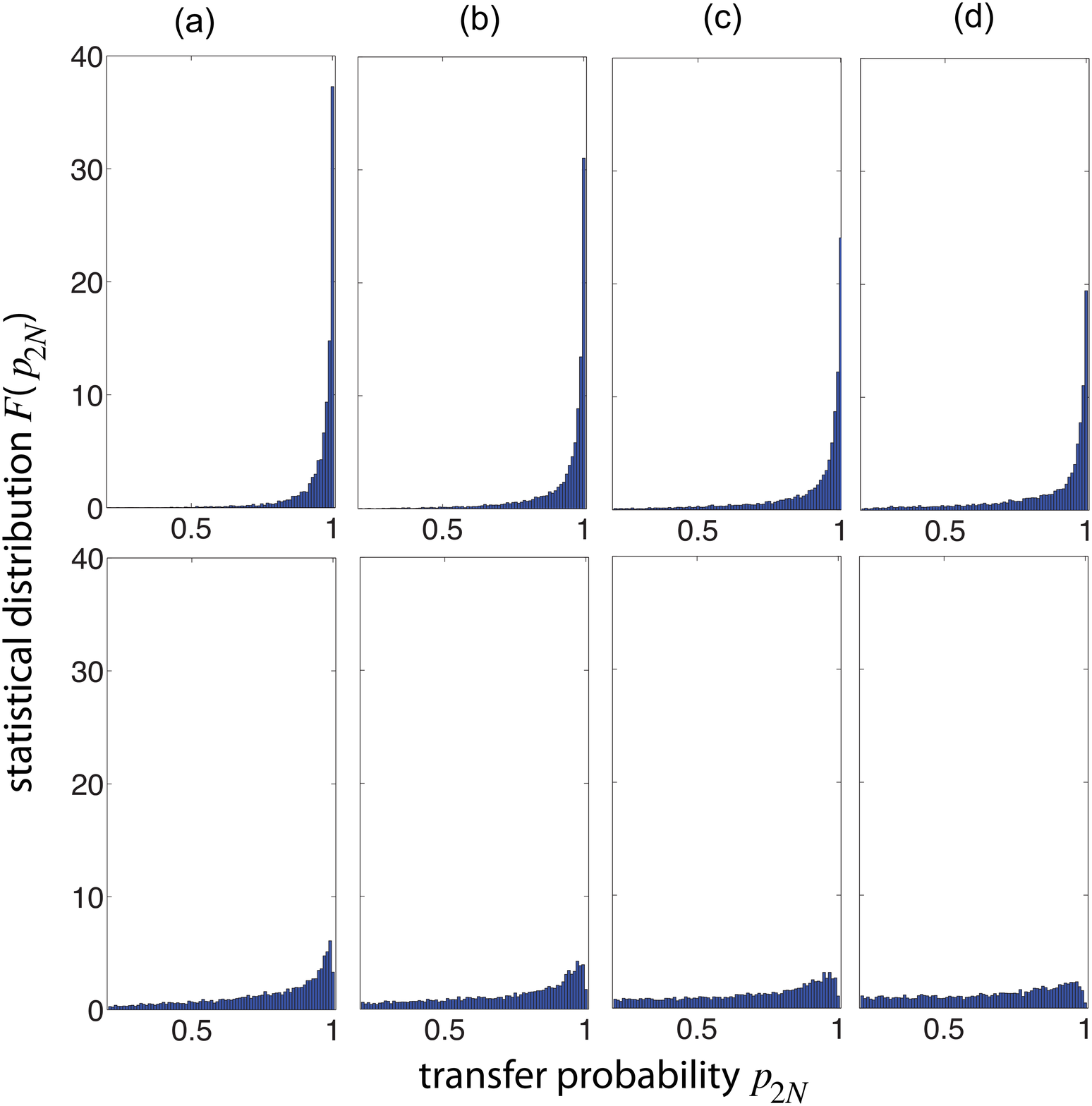}%
  \caption{\label{<label name>}\col 
  {{Statistical distributions of the transfer probability for increasing strength $\delta E$ of diagonal disorder in the adiabatic LZ protocol (upper panels) and in the adiabatic Rabi protocol (lower panels). (a) $\delta E=0.5$, (b) $\delta E=0.6$, (c) $\delta E=0.7$, (d) $\delta E=0.8$. The other parameter values are as in Fig.6 and 7 for the Rabi and LZ protocols, respectively.}}}
\end{figure}
 \section{Effect of disorder on quantum state transfer: comparison between Rabi and Landau-Zener protocols}
 The main advantage of the LZ topological QST method, over Rabi-flopping schemes \cite{r25bis,r27,r27bis}, is to be robust against disorder and structural imperfections of the chain, thus fully harnessing the topological protection feature of edge states. In addition, since the LZ transition is rather insensitive to the precise value of energy splitting of the edge states, the robustness of the LZ QST protocol persists even for disorder that breaks the chiral symmetry of the SSH lattice. We checked that the topological LZ QST scheme is more robust than the adiabatic Rabi flopping scheme by a statistical analysis of the effects of either off-diagonal and on-diagonal disorder on the transfer probability $p_{2N}=|c_{2N}(T)|^2$ in the protocol schemes defined by Eq.(13) (adiabatic Rabi scheme) and Eq.(15) (LZ scheme). The disorder is introduced by considering the modified Hamiltonian $\mathcal{H}+\delta \mathcal{H}$, where $\mathcal{H}$ is the Hamiltonian of the ordered chain given by Eq.(4) and $\delta \mathcal{H}$ accounts for either off-diagonal or on-diagonal disorder. For the sake of simplicity, structural off-diagonal disorder is emulated by introducing random fluctuations of the (static) inter-dimer hopping rate $t_1$ solely around the mean value 1, i.e. we assume
 \begin{equation}
 \mathcal{\delta H}= \left(
 \begin{array}{cccccccccc}
 0 & 0 & 0 & 0 & 0 & ...& 0 & 0 & 0  &0\\
  0 & 0 & \sigma_1 & 0 & 0 & ... & 0 & 0 & 0 & 0 \\
 0 & \sigma_1 & 0 & 0 & 0 &  ... & 0 & 0 & 0  & 0\\
 ... & ... & ...& ...& ... & ...& ...& ...& ...& ...  \\
 0 & 0& 0 & 0& 0 & ...    & 0 & 0 &\sigma_{N-1} & 0 \\
 0 & 0& 0 & 0& 0 & ...    & 0 & \sigma_{N-1} &0 & 0 \\
 0 & 0& 0 & 0& 0 & ...    & 0 & 0 & 0 & 0  \\
 \end{array}
 \right).
 \end{equation} 
 where $\sigma_n$ is a random variable with uniform distribution in the range $(-\sigma,\sigma)$ and $\sigma$ is a measure of the off-diagonal disorder strength. However, we do not expect substantial qualitative changes of results by considering disorder in inter-dimer hopping rate $t_2$ as well, since the main feature of disorder in the SSH lattice is known to arise from the gap closing condition $t_2/t_1=1$ which breaks the topological protection of edge states. Structural on-diagonal disorder is emulated by considering the diagonal Hamiltonian   
  \begin{equation}
 \mathcal{\delta H}= \left(
 \begin{array}{cccccccccc}
 \delta E_1 & 0 & 0 & 0 & 0 & ...& 0 & 0 & 0  &0\\
  0 & \delta E_2 & 0 & 0 & 0 & ... & 0 & 0 & 0 & 0 \\
 0 & 0 & \delta E_3 & 0 & 0 &  ... & 0 & 0 & 0  & 0\\
 ... & ... & ...& ...& ... & ...& ...& ...& ...& ...  \\
 0 & 0& 0 & 0& 0 & ...    & 0 & \delta E_{2N-2} &0 & 0 \\
 0 & 0& 0 & 0& 0 & ...    & 0 & 0 & \delta E_{2N-1}& 0 \\
 0 & 0& 0 & 0& 0 & ...    & 0 & 0 & 0 & \delta E_{2N}  \\
 \end{array}
 \right).
 \end{equation} 
 where $\delta E_n$ is a random variable with uniform distribution in the range $(-\delta E,\delta E)$ and $\delta E$ measures the strength of on-diagonal (site energy) disorder. Statistical analysis has been performed by numerical computation of the transfer excitation probability $p_{2N}$ {{, using the exact Schr\"odinger equation (3),}}  for 10000 realizations of disorder. For the adiabatic Rabi protocol, parameter values used in the simulations are $\epsilon=0.1$ and $T=86$, corresponding to $p_{2N} \simeq 0.995$ in the absence of disorder [see Fig.2(c)]. Figure 6 shows the statistical distribution $F(p_{2N})$ of $p_{2N}$ in the presence of diagonal [Fig.6(a)] and off-diagonal [Fig.6(b)] disorder of moderate strength ($20\%$ in units of the hopping rate $t_1$).  The normalization condition $\int_0^1 dp_{2N} F(p_{2N})=1$ is assumed for the statistical density distribution function $F$. For both diagonal and off-diagonal disorder, $F$ shows a long tail departing from $p_{2N}=1$, indicating that the fidelity of the QST is heavily degraded by structural disorder in the chain, especially in case of diagonal disorder. Such results should be compared to the ones shown in Fig.7, which refer to the impact of the same strength of disorder in the topological LZ protocol. In this case parameter values used in the simulations are those in Fig.3(c) [$\epsilon=0.1$, $\delta_0=0.2$, $\tau=60$, $\tau_Z=120$], corresponding to $p_{2N} \simeq 0.995$ in the absence of disorder. Clearly, in this case the statistical distribution $F$ is much more squeezed toward $p_{2N}=1$, with negligible tails below $p_{2N}=0.9$, indicating that the fidelity of state transfer is not appreciably degraded even in the presence of a moderate disorder in the chain. {{An inspection of Figs. 6 and 7 shows that the diagonal (on-site) disorder is more detrimental than off-diagonal disorder. What happens if we increase the disorder strength further? Clearly,  as the strength of disorder is increased, the transfer probability is degraded in both protocols, however the largest strength of on-diagonal disorder that is tolerated by the LZ protocol is much larger than the one of the Rabi protocol. This is shown in Fig.8, where we compare the statistical distribution $F(p_{2N})$ of the transfer probability $p_{2N}$ for the two protocols for a few increasing values of the diagonal disorder strength $\delta E$. Clearly, even for extremely strong disorder of on-site potential, larger than the staggered magnetic field amplitude $\delta_0$, the LZ protocol shows a strong robustness against disorder, while the Rabi protocol becomes fully unreliable (compare upper and lower panels in Fig.8). This result can be physically explained as follows. In the Rabi protocol, the on-site disorder changes the energy splitting of the edge states in a rather random fashion, so that for a fixed interaction time $T$ the excitation transfer between the two edge sites undergoes large fluctuations because the area on the left hand side of Eq.(14) can greatly deviate from the target value $\pi/2$. In the LZ protocol, the splitting of the edge states also undergoes the same random fluctuation, depending on the precise realization of disorder, however the transfer probability is now much less sensitive to the fluctuations provided that these remain smaller than the amplitude $\delta_0$ of the staggered magnetic field: in fact, in this case the ramp of the magnetic field in stage II of Fig.3(a) will always set the two edge states in resonance and thus LZ tunneling will occur.}}
 
 \section{Conclusions}
 In recent years, topological protection has emerged as a promising route for guiding and transmitting quantum information reliably. Adiabatic (Thouless) pumping of topological states offers some topological protection of quantum state transfer against sizable imperfections in the system \cite{r12bis,r23,r28}.  However, the existence of topological states in a network does not itself ensure that any QST protocol fully exploits the topological protection of states. For example, some recent QST methods based on static or adiabatic Rabi flopping of edge states \cite{r25bis,r27,r27bis} turn out to be sensitive to structural imperfections of the network and thus they require special disorder-dependent timing for the realization of high-fidelity QST. 
In this work we introduced a novel scheme for robust QST of topologically protected edge states in a dimeric Su-Schrieffer-Heeger spin chain assisted by Landau-Zener tunneling. As compared to topological QST protocols based on Rabi flopping, our scheme {{is more advantageous}} in terms of robustness against both diagonal and off-diagonal disorder in the chain, without a substantial increase of  the interaction time.

Our model could be of potential relevance for experimental implementation using  current technology in different setups: possible candidates are chains of superconducting  qubits or optical waveguide lattices. The underlying concepts of our protocol also suggest that topological protection could be exploited in more complicated quantum information tasks, as, for instance, entanglement transfer in structured networks or reservoir engineering.\\ 
\\
{\bf{Acknowledgments.}} G.L.G. acknowledges financial support from the "Consellaria d'Innovació, Recerca i Turisme del Govern de les Illes Balears". S.L. acknowledges hospitality from IFISC-UIB (Palma de Mallorca) under the "professors convidats" program. This work was supported by 
MINECO/AEI/FEDER through project EPheQuCS FIS2016-78010-P.\\
\\
{\bf{Conflict of Interests.}} The authors declare no conflict of interest.\\
\\
{\bf{Keywords.}} Quantum state transfer, spin chains, topological protection.\\



  \appendix
  \renewcommand{\theequation}{A.\arabic{equation}}
\setcounter{equation}{0}
  \section{Reduced two-level model of state transfer dynamics}
In this Appendix we briefly derive the approximate two-level model describing excitation transfer between left $|L \rangle$ and right $|R \rangle$ topological edge states of the SSH chain. The two edge states are defined by Eqs.(5) and (6) given in the main text. For a matrix Hamiltonian $\mathcal{H}$ [Eq.(4)] with constant parameters $t_1$. $t_2$ and $\delta$, it can be readily shown that, in the $N \rightarrow \infty$ limit, $|L \rangle$ and $|R \rangle$ states are eigenstates of $\mathcal{H}$ with eigen-energies $\delta$ and $-\delta$, respectively, i.e.  $\mathcal{H}|L \rangle=\delta |L \rangle$ and $\mathcal{H}|R \rangle=-\delta |R \rangle$. An approximate description of the excitation transfer protocols, which captures the main qualitative features of the process, can be gained by making the rather crude assumption that the dynamics occurs in the subspace of the instantaneous eigenvectors $|L \rangle$ and right $|R \rangle$ of $\mathcal{H}(t)$ (two-level approximation). Such an assumption is a reasonable one provided that (i) the initial excitation state $|\psi(0) \rangle$ is limited to  the two-level subspace (in our case, since $c_n(0)=\delta_{n,1}$, this means $r(0) \equiv t_2(0)/t_1(0) \ll 1$; (ii) the time variation of parameters $t_1$, $t_2$ and $\delta$ is sufficiently slow to neglect non-adiabatic effects; (iii) at each time, the instantaneous localization length $\Lambda$ of edge modes [Eq.(7)] remains smaller than the chain size $N$. We stress that we use the two-level approximation in order to catch the main qualitative features of the transfer dynamics, however it is clear that such a rather crude approximation may fail to provide the exact quantitative analysis of the dynamics, such as the optimal transfer time $T$ and fidelity, which should be computed by numerically solving the Schr\"odinger equation (3) in the full Hilbert space. In particular, the two-level approximation is expected to get less accurate when $r$ gets close to one, i.e. near the gap closing regime, owing to non-adiabatic excitation of bulk states. In the spirit of the two-level approximation, we make the Ansatz
  \begin{equation}
  | \psi(t) \rangle \simeq a_L(t) |L \rangle+ a_R(t) |R \rangle
  \end{equation}
where $a_L(t)$ and $a_R(t)$ are the occupation amplitudes of the two edge states at time $t$. The evolution equations of  $a_{L,R}(t)$ are obtained after substitution of the Anstaz (A.1) into the Schr\"odinger equation $(i d | \psi \rangle /dt)= \mathcal{H} | \psi(t) \rangle$ and multiplying the equation so obtained by $\langle L|$ and $\langle R|$. Taking into account that 
\[
\langle L| R \rangle=\langle L | (d R/dt) \rangle= \langle R | (d L /dt) \rangle=0 \]
and $\langle L | dL/dt \rangle=\langle R | dR/dt \rangle \neq 0$, after gauging out an inessential phase term one obtains
\begin{eqnarray}
i \frac{da_L}{dt} & = & \langle L | \mathcal{H} |L \rangle a_L+\langle L | \mathcal{H} |R \rangle a_R  =  \delta a_L+\kappa a_R \\
i \frac{da_R}{dt} & = & \langle R | \mathcal{H} |L \rangle a_L+\langle R | \mathcal{H} |R \rangle a_R  = - \delta a_R+\kappa a_L 
\end{eqnarray}
where $\kappa$ is given by
\begin{equation}
\kappa  \equiv  \langle L | \mathcal{H} |R \rangle=\langle R | \mathcal{H} |L \rangle=\frac{t_1 \left( t_2/t_1\right)^{N} \left[ (t_2/t_1)^2-1\right]}{(t_2/t_1)^{2N}-1}.
\end{equation}


\begin{thebibliography}{0}
  
\bibitem{r1}
P. Kral, I. Thanopulos,  M. Shapiro, \textit{Rev. Mod. Phys.} \textbf{2007},  \textit{79}, 53.
\bibitem{r2}
H. Dong, D.-Z. Xu, J.-F. Huang,  C.-P. Sun, \textit{Light: Science \& Applications} \textbf{2012}, \textit{1}, e2.
\bibitem{r3}
 A. Thilagam, \textit{J. Chem. Phys.} \textbf{2012}, \textit{136}, 065104.
 \bibitem{r4}
V. Abramavicius, V. Pranculis, A. Melianas, O. Ingan\"as, V. Gulbinas, D. Abramavicius,\textit{ Sci. Rep.} \textbf{2016}, \textit{6}, 32914.
\bibitem{r5}
S. Bose, \textit{Phys. Rev. Lett.} \textbf{2003}, \textit{91}, 20790.
\bibitem{r6}
 S. Bose, \textit{Contemp. Phys.} \textbf{2007},  \textit{48}, 13.
\bibitem{r7}
A. Kay, \textit{Int. J. Quantum Inf.} \textbf{2010},  \textit{8},  641.
\bibitem{r8}
 A. Kay, \textit{Phys. Rev. A} \textbf{2009},  \textit{79}, 042330.
\bibitem{r9} 
G.M. Nikolopoulos, I. Jex, {\it Quantum State Transfer and Network Engineering}, Springer-Verlag, Berlin, Germany, \textbf{2014}.
\bibitem{r10}
G.M. Nikolopoulos, D. Petrosyan, P. Lambropoulos, \textit{EPL} \textbf{2004}, \textit{ 65}, 297.
\bibitem{r11}
M. Christandl, N. Datta, A. Ekert, A.J. Landahl, \textit{Phys. Rev. Lett.} \textbf{2004}, \textit{92},  187902.
\bibitem{r12}
 M.B. Plenio, J. Hartley,  J. Eisert,\textit{ New J. Phys.} \textbf{2004},  \textit{6}, 36.
 \bibitem{r12bis}
 K Eckert, O. Romero-Isart,  A. Sanpera, \textit{New J. Phys.} \textbf{2007},  \textit{9}, 155.
\bibitem{r13}
R. Menchon-Enrich, A. Benseny, V. Ahufinger, A.D. Greentree, T. Busch,  J. Mompart, \textit{Rep. Prog. Phys.} \textbf{2016}, \textit{79}, 074401.
\bibitem{r14}
C.E. Creffield, \textit{Phys. Rev. Lett.} \textbf{2007}, \textit{99}, 110501.
\bibitem{r15}
M.X. Huo, Y. Li, Z. Song,  C.P. Sun, \textit{EPL} \textbf{2008}, \textit{84}, 30004.
\bibitem{r16}
N.Y. Yao, L. Jiang, A.V. Gorshkov, Z.-X. Gong, A. Zhai, L.-M. Duan,  M. D. Lukin, \textit{Phys. Rev. Lett.} \textbf{2011},  \textit{106}, 040505.
\bibitem{r17}
S. Paganelli, S. Lorenzo, T.J.G. Apollaro, F. Plastina,  G.L. Giorgi,
\textit{Phys. Rev. A} \textbf{2013}, \textit{87}, 062309.
\bibitem{r18}
S. Lorenzo, T.J.G. Apollaro, A. Sindona, F. Plastina,
\textit{Phys. Rev. A} \textbf{2013}, \textit{87}, 042313.
\bibitem{r19}
S. Lorenzo, T.J.G. Apollaro, S. Paganelli, G.M. Palma,  F. Plastina, \textit{Phys. Rev. A } \textbf{2015}, \textit{91}, 042321.
\bibitem{r19bis}
S. Longhi, \textit{EPL} \textbf{2016}, \textit{113}, 60006.
\bibitem{r19tris}
M.P. Silveri, K.S. Kumar, J. Tuorila, J. Li, A. Veps\"al\"ainen, E.V. Thuneberg, G.S. Paraoanu, 
\textit{New J. Phys.} \textbf{2015}, \textit{17}, 043058.
\bibitem{r20}
M. Bellec, G.M. Nikolopoulos,  S. Tzortzakis, \textit{Opt.
Lett.} \textbf{2012}, \textit{37}, 4504.
\bibitem{r21}
A. Perez-Leija, R. Keil, A. Kay,
H. Moya-Cessa, S. Nolte, L.-C. Kwek, B.M. Rodriguez-
Lara, A. Szameit,  D.N. Christodoulides, \textit{Phys. Rev.
A} \textbf{2013},  \textit{87}, 012309.
\bibitem{r22}
R.J. Chapman, M. Santandrea, Z. Huang, G. Corrielli, A. Crespi,
M.-H. Yung, R. Osellame,  A. Peruzzo, \textit{	Nat. Commun.} \textbf{2016}, \textit{7}, 11339. 
\bibitem{r23uff}
G. Della Valle, M. Ornigotti, T. Toney Fernandez, P. Laporta, S. Longhi, A. Coppa,  V. Foglietti, \textit{Appl. Phys. Lett.} \textbf{2008}, \textit{92}, 011106.
\bibitem{r23}
Y.E. Kraus, Y. Lahini, Z. Ringel, M. Verbin,  O. Zilberberg, \textit{Phys. Rev. Lett.} \textbf{2012}, \textit{109}, 106402.
\bibitem{r24}
N.Y. Yao, C.R. Laumann, A.V. Gorshkov, H. Weimer, L. Jiang, J.I. Cirac, P. Zoller,  M.D. Lukin, \textit{	Nat. Commun.} \textbf{2013},  \textit{4}, 1585.
\bibitem{r25}
M. Verbin, O. Zilberberg, Y. Lahini, Y.E. Kraus,  Y. Silberberg, \textit{Phys. Rev. B } \textbf{2015}, \textit{91}, 064201.
\bibitem{r25bis}
M. Bello, C. E. Creffield,  G. Platero, \textit{Sci. Rep.} \textbf{2016}, \textit{6}, 22562.
\bibitem{r25tris}
J.-L. Wu, X. Ji,  S. Zhang,\textit{ Sci. Rep.} \textbf{2017},  \textit{7}, 46255.
\bibitem{r25quatris}
R. Agundez, C.D. Hill, L.C.L. Hollenberg, S. Rogge,  M. Blaauboer, \textit{Phys. Rev. A} \textbf{2017},  \textit{95}, 012317.
\bibitem{r26}
C. Dlaska, B. Vermersch1,  P Zoller, \textit{Quantum Sci. Technol.} \textbf{2017}, \textit{2}, 015001.
\bibitem{r27}
N. Lang and H.P. B\"uchler, \textit{npj Quantum Inf.} \textbf{2017}, \textit{3}, 47.
\bibitem{r27bis}
M.P. Estarellas, I. D'Amico,  T.P. Spiller, \textit{Sci. Rep. } \textbf{2017}, 7, 42904.
\bibitem{r28}
F. Mei, G. Chen, L. Tian, S.-L. Zhu,  S. Jia, \textit{Phys. Rev. A} \textbf{2018}, \textit{98}, 012331.
\bibitem{r29}
W. P. Su, J. R. Schrieffer,  A. J. Heeger, \textit{Phys. Rev. Lett.} \textbf{1979}, \textit{42}, 1698.
\bibitem{uffa1}
S. Longhi, \textit{Laser \& Photon. Rev.} \textbf{2009}, \textit{3}, 243. 
\bibitem{Alex1}
M.C. Rechtsman, J.M. Zeuner, Y. Plotnik, Y. Lumer, D. Podolsky, F. Dreisow, S. Nolte, M. Segev, A. Szameit, \textit{Nature} \textbf{2013}, \textit{496}, 196.
\bibitem{Alex2}
M.C. Rechtsman, Y. Plotnik, J.M. Zeuner, D. Song, Z. Chen, A. Szameit, M. Segev, \textit{Phys. Rev. Lett.} \textbf{2013}, \textit{111}, 103901. 
\bibitem{Alex3}
A. Blanco-Redondo, B. Bell,  D. Oren, B.J. Eggleton, M. Segev, \textit{Science} \textbf{2018}, \textit{362}, 568.
\bibitem{r34bis}
J.-L. Tambasco, G. Corrielli, R.J. Chapman, A. Crespi, O. Zilberberg, R. Osellame,  A. Peruzzo, \textit{Sci. Adv.} \textbf{2018},  \textit{4}, eaat3187.
\bibitem{r31}
G.L. Giorgi, \textit{Phys. Rev. B} \textbf{2009},  \textit{79}, 060405(R).
\bibitem{r37bis}
M. J. Rice and E. J. Mele, \textit{Phys. Rev. Lett.} \textbf{1982}, \textit{49}, 1455.
\bibitem{r30}
N.V. Vitanov, T. Halfmann, B.W. Shore,  K. Bergmann, \textit{Ann. Rev. Phys. Chem.} \textbf{2001},  \textit{52}, 763.
\bibitem{r30a}
L.F. Wei, J.R. Johansson, L.X. Cen, S. Ashhab,  F. Nori, \textit{ Phys. Rev. Lett.} \textbf{2008}, \textit{100}, 113601.
\bibitem{r30b}
S. Shevchenko, S. Ashhab,  F. Nori, \textit{Phys. Rep.} \textbf{2010}, \textit{492}, 1.
\bibitem{r30c}
G. Sun, X. Wen, B. Mao, J. Chen, Y. Yu, P. Wu,  S. Han, \textit{Nat. Commun.} \textbf{2010}, \textit{1}, 51.
\bibitem{r30d}
S. Gasparinetti, P. Solinas,  J. P. Pekola, \textit{Phys. Rev. Lett.} \textbf{2011}, \textit{107}, 207002.
\bibitem{r30e}
X. Tan, D.-W. Zhang, Z. Zhang, Y. Yu, S. Han, S.-L. Zhu, \textit{Phys. Rev. Lett.} \textbf{2014}, \textit{112}, 027001.
\bibitem{r30f}
A. Ferron, D. Domínguez, M. J. Sanchez, \textit{Phys. Rev. B} \textbf{2016}, \textit{93}, 064521.
\bibitem{r32}
S. Longhi and G. Della Valle, \textit{Phys. Rev. A} \textbf{2012}, \textit{86}, 043633.
\bibitem{palle1}
W.D. Oliver, Y. Yu, J.C. Lee, K.K. Berggren, L.S. Levitov, T.P. Orlando, \textit{Science} \textbf{2005}, \textit{310}, 1653.
\bibitem{palle2}
M. Sillanp\"a\"a, T. Lehtinen, A. Paila, Y. Makhlin, P. Hakonen, \textit{Phys. Rev. Lett.} \textbf{2006}, \textit{96}, 187002.
\bibitem{palle3}
X. Li, Y. Ma, J. Han, Tao Chen, Y. Xu, W. Cai, H. Wang, Y. P. Song, Z.-Y. Xue, Z.-q. Yin, L. Sun,
\textit{Phys. Rev. Applied} \textbf{2018}, \textit{10}, 054009.
\bibitem{r33}
B. Chen, Y.-D. Peng, Y. Li,  X.-F. Qian, \textit{Sci. Rep.}  \textbf{2016}, \textit{6}, 28886.
\bibitem{Robinson}
E.H. Lieb, D.W. Robinson,  \textit{Commun. Math. Phys.} \textbf{1972}, \textit{28}, 251.

\end{thebibliography}
\end{document}